\documentclass[12pt]{article}

\renewcommand{\baselinestretch}{1.5}

\newcommand{\sth}{\sigma^3}
\newcommand{\stw}{\sigma^2}

\newcommand{\sfo}{\sigma^4}
\newcommand{\sfoprime}{\sigma^{4\prime}}
\newcommand{\alstwsfo}{\alpha_{\sigma^2\sigma^4}}
\newcommand{\Astwsfo}{A_{\sigma^2\sigma^4}}

\newcommand{\Astw}{A_{\sigma^2}}
\newcommand{\be}{\mbox{\boldmath$e$}}
\newcommand{\bvarphi}{\mbox{\boldmath$\varphi$}}
\newcommand{\bpsi}{\mbox{\boldmath$\psi$}}
\newcommand{\sumsfosupstw}{\sum_{\sigma^4 \supset \sigma^2}}
\newcommand{\sumstwsubsfo}{\sum_{\sigma^2 \subset \sigma^4}}
\newcommand{\sumalpha}{\sum^4_{\alpha = 1}}
\newcommand{\sumbeta}{\sum^4_{\beta = 1}}
\newcommand{\sumalphabeta}{\sum^4_{\alpha < \beta}}
\newcommand{\sumstw}{\sum_{\sigma^2}}
\newcommand{\sumsfo}{\sum_{\sigma^4}}
\newcommand{\bisim}{\triangle\!\!\!\nabla}

\newcommand{\pmr}{\,^{\pm}\!r}
\newcommand{\pmbralpha}{\,^{\pm}\!\mbox{\boldmath$r$}_{\alpha}}
\newcommand{\pbralpha}{\,^{+}\!\mbox{\boldmath$r$}_{\alpha}}
\newcommand{\mbralpha}{\,^{-}\!\mbox{\boldmath$r$}_{\alpha}}

\newcommand{\pmbn}{\,^{\pm}\!\mbox{\boldmath$n$}}
\newcommand{\pmbe}{\,^{\pm}\!\mbox{\boldmath$e$}}
\newcommand{\pmbm}{\,^{\pm}\!\mbox{\boldmath$m$}}
\newcommand{\pbm}{\,^{+}\!\mbox{\boldmath$m$}}
\newcommand{\mbm}{\,^{-}\!\mbox{\boldmath$m$}}
\newcommand{\pmm}{\,^{\pm}\!m}
\newcommand{\pmn}{\,^{\pm}\!n}
\newcommand{\pbr}{\,^{+}\!\mbox{\boldmath$r$}}
\newcommand{\mbr}{\,^{-}\!\mbox{\boldmath$r$}}

\newcommand{\pbrbeta}{\,^{+}\!\mbox{\boldmath$r$}_{\beta}}

\newcommand{\Nstw}{N_{\sigma^2}}
\newcommand{\N}{{\cal N}}
\newcommand{\pmcalN}{\,^{\pm}\!{\cal N}}
\newcommand{\pcalN}{\,^{+}\!{\cal N}}
\newcommand{\mcalN}{\,^{-}\!{\cal N}}
\newcommand{\Sbisimsfo}{S_{\triangle\!\!\!\nabla}(\sigma^4 )}
\newcommand{\Omsth}{\Omega_{\sigma^3}}
\newcommand{\Omsthsfo}{\Omega_{\sigma^3\sigma^4}}

\newcommand{\Rstw}{R_{\sigma^2}}
\newcommand{\Rstwsfo}{R_{\sigma^2\sigma^4}}
\newcommand{\Arcsin}{{\rm Arcsin}}
\newcommand{\plS}{\,^+\!S}
\newcommand{\miS}{\,^-\!S}
\newcommand{\plm}{\,^+\!m}
\newcommand{\mim}{\,^-\!m}
\newcommand{\pmS}{\,^{\pm}\!S}
\newcommand{\pmSig}{\,^{\pm}\!\Sigma}
\newcommand{\pmbSig}{\,^{\pm}\!\mbox{\boldmath$\Sigma$}}
\newcommand{\pmSSUtwbisim}{\,^{\pm}\!S_{\triangle\!\!\!\nabla}^{\rm SU(2)}}
\newcommand{\pmSSOthbisim}{\,^{\pm}\!S_{\triangle\!\!\!\nabla}^{\rm SO(3)}}
\newcommand{\pmbvstwsfo}{\,^{\pm}\!\mbox{\boldmath$v$}_{\sigma^2\sigma^4}}
\newcommand{\pmbvalpha}{\,^{\pm}\!\mbox{\boldmath$v$}_{\alpha}}
\newcommand{\pbvalpha}{\,^{+}\!\mbox{\boldmath$v$}_{\alpha}}
\newcommand{\mbvalpha}{\,^{-}\!\mbox{\boldmath$v$}_{\alpha}}
\newcommand{\pmbvalphabeta}{\,^{\pm}\!\mbox{\boldmath$v$}_{\alpha\beta}}
\newcommand{\pmbvbetaalpha}{\,^{\pm}\!\mbox{\boldmath$v$}_{\beta\alpha}}
\newcommand{\pbvbetaalpha}{\,^{+}\!\mbox{\boldmath$v$}_{\beta\alpha}}
\newcommand{\mbvbetaalpha}{\,^{-}\!\mbox{\boldmath$v$}_{\beta\alpha}}
\newcommand{\pbvalphabeta}{\,^{+}\!\mbox{\boldmath$v$}_{\alpha\beta}}
\newcommand{\pmvstwsfo}{\,^{\pm}\!v_{\sigma^2\sigma^4}}
\newcommand{\pmvalpha}{\,^{\pm}\!v_{\alpha}}
\newcommand{\pvalpha}{\,^{+}\!v_{\alpha}}
\newcommand{\mvalpha}{\,^{-}\!v_{\alpha}}
\newcommand{\pkalpha}{\,^{+}\!k_{\alpha}}
\newcommand{\mkalpha}{\,^{-}\!k_{\alpha}}
\newcommand{\pmvalphabeta}{\,^{\pm}\!v_{\alpha \beta}}
\newcommand{\bvalpha}{\mbox{\boldmath$v$}_{\alpha}}
\newcommand{\valpha}{v_{\alpha}}
\newcommand{\valphabeta}{v_{\alpha \beta}}
\newcommand{\pmRstwsfo}{\,^{\pm}\!R_{\sigma^2\sigma^4}}
\newcommand{\pmv}{\,^{\pm}\!v}
\newcommand{\pmtilv}{\,^{\pm}\!\tilde{v}}
\newcommand{\ptilv}{\,^{+}\!\tilde{v}}

\newcommand{\pmbvstw}{\,^{\pm}\!\mbox{\boldmath$v$}_{\sigma^2}}
\newcommand{\bv}{\mbox{\boldmath$v$}}
\newcommand{\pmbv}{\,^{\pm}\!\mbox{\boldmath$v$}}
\newcommand{\pbv}{\,^{+}\!\mbox{\boldmath$v$}}
\newcommand{\mbv}{\,^{-}\!\mbox{\boldmath$v$}}

\newcommand{\pmR}{\,^{\pm}\!R}
\newcommand{\pmRT}{\,^{\pm}\!R^{\rm T}}
\newcommand{\pR}{\,^{+}\!R}
\newcommand{\pRT}{\,^{+}\!R^{\rm T}}
\newcommand{\mR}{\,^{-}\!R}
\newcommand{\mRT}{\,^{-}\!R^{\rm T}}
\newcommand{\pmbphi}{\,^{\pm}\!\mbox{\boldmath$\phi$}}
\newcommand{\pmphi}{\,^{\pm}\!\phi}
\newcommand{\pbphi}{\,^{+}\!\mbox{\boldmath$\phi$}}
\newcommand{\mbphi}{\,^{-}\!\mbox{\boldmath$\phi$}}
\newcommand{\pphi}{\,^{+}\!\phi}
\newcommand{\mphi}{\,^{-}\!\phi}
\newcommand{\pOmega}{\,^+\!\Omega}
\newcommand{\mOmega}{\,^-\!\Omega}
\newcommand{\pmOmega}{\,^{\pm}\!\Omega}
\newcommand{\OmT}{\Omega^{\rm T}}
\newcommand{\RT}{R^{\rm T}}

\newcommand{\prodsfo}{\prod_{\sigma^4}}
\newcommand{\prodsthsfo}{\prod_{\sigma^3 , \sigma^4}}
\newcommand{\prodsthsubsfo}{\prod_{\sigma^3 \subset \sigma^4}}
\newcommand{\prodalpha}{\prod^4_{\alpha = 1}}
\newcommand{\pv}{\,^{+}\!v}
\newcommand{\mv}{\,^{-}\!v}
\newcommand{\bl}{\mbox{\boldmath$l$}}

\newcommand{\sumsthsubsfo}{\sum_{\sigma^3 \subset \sigma^4}}
\newcommand{\pmSSOth}{\,^{\pm}\!S^{\rm SO(3)}}
\newcommand{\pmSSUtw}{\,^{\pm}\!S^{\rm SU(2)}}

\renewcommand{\d}{{\rm d}}
\renewcommand{\Re}{{\mbox{Re}}}
\renewcommand{\Im}{{\mbox{Im}}}

\newcommand{\dfun}{$\delta$-function }
\newcommand{\dfuns}{$\delta$-functions }
\newcommand{\desi}{\delta^6}
\newcommand{\deth}{\delta^3}

\newcommand{\lfoze}{l_{40}}
\newcommand{\lfoalpha}{l_{4 \alpha}}
\newcommand{\lfoon}{l_{41}}
\newcommand{\lfotw}{l_{42}}
\newcommand{\lfoth}{l_{43}}

\newcommand{\blfoalpha}{\mbox{\boldmath$l$}_{4\alpha}}
\newcommand{\blfoon}{\mbox{\boldmath$l$}_{41}}
\newcommand{\blfotw}{\mbox{\boldmath$l$}_{42}}
\newcommand{\blfoth}{\mbox{\boldmath$l$}_{43}}

\newcommand{\D}{{\cal D}}

\newcommand{\pmbeta}{\,^{\pm}\!\beta}
\newcommand{\pmeta}{\,^{\pm}\!\eta}
\newcommand{\peta}{\,^+\!\eta}
\newcommand{\meta}{\,^-\!\eta}
\newcommand{\pmtheta}{\,^{\pm}\!\theta}%1
\newcommand{\pmzeta}{\,^{\pm}\!\zeta}
\newcommand{\pzeta}{\,^+\!\zeta}
\newcommand{\mzeta}{\,^-\!\zeta}
\newcommand{\pmchi}{\,^{\pm}\!\chi}
\newcommand{\pchi}{\,^+\!\chi}%1
\newcommand{\mchi}{\,^-\!\chi}%1
\newcommand{\sh}{{\rm sh}}
\newcommand{\ch}{{\rm ch}}

\begin{document}

\title{A version of the connection representation of Regge action}
\author{V.M. Khatsymovsky \\
 {\em Budker Institute of Nuclear Physics} \\ {\em
 Novosibirsk,
 630090,
 Russia}
\\ {\em E-mail address: khatsym@inp.nsk.su}}
\date{}
\maketitle
\begin{abstract}
We define for any 4-tetrahedron (4-simplex) the simplest finite closed piecewise flat manifold consisting of this 4-tetrahedron and of the one else 4-tetrahed\-ron identical up to reflection to the present one (call it bisimplex built on the given 4-simplex, or two-sided 4-simplex). We consider arbitrary piecewise flat manifold. Gravity action for it can be expressed in terms of sum of the actions for the bisimplices built on the 4-simplices constituting this manifold. We use representation of each bisimplex action in terms of rotation matrices (connections) and area tensors. This gives some representation of any piecewise flat gravity action in terms of connections. The action is a sum of terms each depending on the connection variables referring to a single 4-tetrahedron. Application of this representation to the path integral formalism is considered. Integrations over connections in the path integral reduce to independent integrations over finite sets of connections on separate 4-simplices. One of the consequences is exponential suppression of the result at large areas or lengths (compared to Plank scale). It is important for the consistency of the simplicial description of spacetime.

\end{abstract}


%PACS numbers: 04.60.-m Quantum gravity

%PACS numbers: 31.15.xk Path-integral methods; 11.15.Ha Lattice gauge theory; 04.60.Kz %Minisuperspace models

%MSC classes: 83C27 Lattice gravity, Regge calculus and other discrete methods; 53C05 %Connections, general theory; 81S40 Path integrals

\newpage

\vspace{-7mm}

\section{Introduction}

\vspace{-3mm}

Recently one often considers a modification of the genuine Regge calculus (RC) \cite{Regge} where the same edge can have different lengths depending on the 4-simplex where it is defined, e. g., the so-called area RC \cite{BarRocWil,RegWil}. If we invoke description of the RC in terms of tetrad and connection \cite{Fro}, it is natural to study analogous modification in the connection sector. Can connection on the 3-simplex depend also on the 4-simplex containing it? If variables are divided into independent sets referred to separate 4-simplices, the theory is more simple. An idea is to apply connection representation separately to the (properly specified) contributions to the action of the different 4-tetrahedra.

An idea of how to specify contribution from separate 4-simplices admitting the connection representation can be illustrated by the 2-dimensional case. Consider 2-simplex (triangle) and closed (strongly curved) surface consisting of both sides of this triangle viewed as different 2-simplices. Call it (two-dimensional) {\it bisimplex}. Evidently, the angles of the triangle can be expressed in terms of the defect angles of the bisimplex. Analogously, hyperdihedral angles of any 4-tetrahedron can be expressed in terms of the defect angles of the bisimplex consisting of only two 4-tetrahedra with mutually identified vertices. These can be viewed as different sides of the same 4-tetrahedron. The Regge action of arbitrary simplicial spacetime is combination of actions of the bisimplices constructed by this method from separate 4-simplices. The action of the bisimplex can be written in the connection representation using some connection orthogonal rotational matrices as independent variables. This representation is considered in Section {\bf \ref{bisimplex}}. The representation of the full gravity action is then obtained in Section {\bf \ref{full-action}} as linear combination of the representations of the actions for bisimplices built on separate 4-simplices.

The connection variables are generally arbitrary orthogonal matrices, but on the equations of motion the curvature matrices formed of these should have physical sense of rotations in the local frame of the 4-simplex around its two-dimensional faces (triangles) by hyperdihedral angles of the 4-simplex at these triangles (more accurately, by $2 \pi - 2 \cdot \mbox{(hyperdihedral angle)}$). There is no need in the explicit presence of the above bisimplices in general simplicial complex. Sufficient is that the Regge action for the general simplicial spacetime consists of the terms which can be interpreted as actions for bisimplices built on separate 4-simplices. Therefore all these terms and thus this Regge action are represented using new variables having possible geometrical sense of rotations by hyperdihedral angles. The sense of this representation is that on the equations of motion for the rotational variables it is just the Regge action in terms of the purely edge lengths.

The use of the representation studied is shown in the path integral formalism in Section {\bf \ref{pathint}}. Integrations over connections reduce to independent integrations over finite sets of connections on separate 4-simplices and are analyzable. One of the consequences is exponential suppression of the path integral at large areas or lengths. This means suppression of the physical amplitudes with large areas/lengthgs. This is important for the consistency of the simplicial minisuperspace description of spacetime which should be close to the continuum one at large scales.

The connection representation includes '$\arcsin$' functions which serve to express the hyperdihedral angles which normally are not small. Therefore it is necessary to define proper branches of the '$\arcsin$' functions. This implies division of the whole region of variation of the edge lengths into certain sectors. Section {\bf \ref{branches}} considers typical example of the simplicial structure. Considered is the region of variation of the edge lengths which contains zero curvature physical configuration as a particular case. Analytical form of the representation studied is specified and expressed in terms of combination of the principal value of '$\arcsin$' functions together with terms which are constants different in the different sectors. These terms are important for possible integration over edge lengths in the path integral and do not influence on the integrations over connections of Section {\bf \ref{pathint}}.

\vspace{-7mm}

\section{Bisimplex}\label{bisimplex}

\vspace{-3mm}

Given 4-simplex $\sfo$, consider the above mentioned bisimplex, simplest simplicial complex $\bisim$ built of only two 4-simplices, the vertices of which are mutually identified, $\sfo$ and identical to it, up to reflection w.r.t. any 3-face, 4-simplex $\sfoprime$. Let us write Regge action for the bisimplex constructed of the given 4-simplex $\sfo$,
\begin{equation}\label{Sbisim}
S_{\bisim}(\sfo ) = \sumstwsubsfo (2\pi - 2\alstwsfo )\Astwsfo.
\end{equation}

\noindent Here $\alstwsfo$ is hyperdihedral angle of the 4-simplex $\sfo$ at the 2-face $\stw$, $\Astwsfo$ is the area (either real or imaginary) of the triangle $\stw$ in the 4-simplex $\sfo$.

Invoking the notion of discrete tetrad and connection first considered in Ref. \cite{Fro} we have suggested in Ref. \cite{Kha} representation of the minisuperspace Regge action in terms of area tensors and finite rotation SO(4) (SO(3,1) in the Minkowsky case) matrices, and also in terms of (anti-)selfdual parts of finite rotation matrices. The latter is based on the decomposition of SO(4) as SU(2) $\!\times\!$ SU(2) (more accurately, modulo the element (-1,-1)) or SO(3) $\!\times\!$ SO(3) (more accurately, plus the same times the element -1 of SO(4)). Then SO(3,1) element is representable by two mutually complex conjugated elements of SU(2) or SO(3) with complex parameters (that is, of SU(2,C) or SO(3,C)). For definiteness, one might imply the following notations and sign conventions concerning splitting the tensors into (anti-)selfdual parts in the Minkowsky spacetime. Suppose there is SO(3,1) matrix,
\begin{equation}
\Omega = \exp {(\varphi^kE^a_{kb} + \psi^kL^a_{kb})}. %\nonumber
\end{equation}
Its generator is expanded over the set of independent generators,
\begin{equation}
E_{kab} = -\epsilon_{kab}, ~~~ L_{kab} = g_{ka}g_{0b} - g_{0a}g_{kb}~~~(g_{ab} =
{\rm diag}(-1,1,1,1), \epsilon_{123} = +1 ). %\nonumber
\end{equation}
We denote
\begin{equation}
\pmSig_{kab} = -\epsilon_{kab} \pm i(g_{ak}g_{0b} - g_{a0}g_{kb}) %\nonumber
\end{equation}
so that
\begin{equation}
 ^*(\pmSig^{ab}) \! \equiv \frac{1}{2} \epsilon^{ab}_{~~cd}\pmSig^{cd} \! = \mp i\pmSig^{ab} ~ ( \epsilon^{0123} \! = \! +1), ~ \pmSig^a_{kb} \! \pmSig^b_{lc} \! = - \delta_{kl} \delta^a_c \! + \epsilon_{kl}^{~~m}
\pmSig^a_{mc}, %\nonumber
\end{equation}
then
\begin{equation}\label{varphi-mp-ipsi}
\Omega = \pOmega \mOmega, ~~~ \pmOmega = \exp \left (\frac{\varphi^k \mp i\psi^k} {2}\pmSig^a_{kb} \right ) = \cos \frac{ \!\!\! \pmphi }{2} + \pmbSig \pmbn \sin \frac{ \!\!\! \pmphi }{2}. %\nonumber
\end{equation}
Here $\pmphi = \sqrt{\pmbphi^2}, \pmbphi = \bvarphi \mp \bpsi , \pmbn = \pmbphi / \pmphi$. For a triangle spanned by the two 4-vectors $l^c_1$, $l^d_2$ we can define {\it bivector} $v^{ab} = \frac{1}{2}\epsilon^{ab}_{~~cd}l^c_1l^d_2$ or, as a shorthand, $[l_1, l_2 ]^{ab}$. This variable splits additively,
\begin{equation}
v^{ab} = \pv^{ab} + \mv^{ab}, ~~~ \pmv^{ab} = \frac{1}{2} v^{ab} \pm \frac{i}{4}
\epsilon^{ab}_{~~cd}v^{cd}. %\nonumber
\end{equation}
In particular,
\begin{equation}
2\pmv \circ \pmv = v \circ v \pm iv*v. %\nonumber
\end{equation}
Here $A\circ B \equiv \frac{1}{ 2}A_{ab} B^{ab}$, $A*B$ $\equiv$ $\frac{1}{4}\epsilon_{abcd}A^{ab}B^{cd}$  for the two matrices $A, B$. The $^{\pm}$-parts map into three-dimensional vectors $ \pmbv $,
\begin{equation}\label{v-vec}
\pmv_{ab} \equiv \frac{1}{2}\pmv^k\pmSig_{kab}, ~~~ 2\pmv_k = - \epsilon_{klm}
v^{lm} \pm i(v_{k0} -v_{0k}).
\end{equation}
For a bivector $2 \pmbv = \pm i \bl_1 \times \bl_2 - \bl_1 l^0_2 + \bl_2 l^0_1$. Additional overall $i$ compared to usual definition of the area vector (real for the spacelike area) is here due to the fact that $v^{ab}$ is {\it dual} area tensor. Besides that,
\begin{equation}
\pmbv^2 = 2\pmv \circ \pmv. %\nonumber
\end{equation}
The $\pmbv^2$ is $(-1)$ times the square of the (real for the spacelike triangle) area. The SU(2) representation for $\pmOmega$ (\ref{varphi-mp-ipsi}) means the SO(3) rotation in the space of $ \pmbv $,
\begin{equation}\label{Omega-SO3}
\pmOmega^{ab} = \pmn^a \pmn^b + (\delta^{ab} - \pmn^a \pmn^b ) \cos \pmphi + \epsilon^{abc} \pmn_c \sin \pmphi .
\end{equation}

The considered SU(2) $\!\times\!$ SU(2) representation in terms of (anti-)selfdual parts of finite rotation matrices can be written for the considered manifold as
\begin{eqnarray}
\label{SU2}%
\pmSSUtwbisim (\sfo ) & = & \sumstwsubsfo \sqrt{\pmbvstwsfo^2} \Arcsin {\pmvstwsfo \circ \pmRstwsfo ( \Omega ) \over \sqrt{\pmbvstwsfo^2}}.
\end{eqnarray}

\noindent Here $\pmbvstw$ are area vectors (\ref{v-vec}) of the triangle $\stw$, in the Minkowsky case $\Omsth$ is rotation SO(3,1) matrix on the tetrahedron $\sth$ which we call simply connection, $\Rstw$ is curvature matrix on the triangle $\stw$ (holonomy of $\Omega$'s). The vector/tensor indices of $v, R$ might refer to the frame of any one of the two 4-simplices $\sfo$, $\sfoprime$; for definiteness we choose the original $\sfo$. Further, there are 5 connection matrices $\Omsth$ in our particular simplicial complex and 10 curvature matrices $\Rstwsfo$, each $R$ being product of certain two matrices $\Omega^{\pm 1}$. The '$\Arcsin$' means proper solution for the inverse function to '$\sin$' while '$\arcsin$' means principal value whose real part at real argument lays in the region $[-\pi/2, +\pi/2]$. The superscript SU(2) means that $\pmv$ can be viewed as 2 $\times$ 2 antisymmetric matrix, and $\pmR$ as SU(2) matrix (to be precise, SU(2,C) in Minkowsky case). This is fundamental SU(2) representation. Also adjoint, SO(3) $\!\times\!$ SO(3) representation is of interest for us,
\begin{eqnarray}
\label{SO3}%
\pmSSOthbisim (\sfo ) & = & \sumstwsubsfo {1\over 2} \sqrt{\pmbvstwsfo^2} \Arcsin {\pmbvstwsfo * \pmRstwsfo ( \Omega ) \over \sqrt{\pmbvstwsfo^2}}.
\end{eqnarray}

\noindent Here $\pmRstwsfo$ is SO(3) matrix (to be precise, SO(3,C) in Minkowsky case); for a 3-vector $\bv$ and a $3\times 3$ matrix $R$ we have denoted $\bv * R \equiv {1\over 2}v^a R^{bc} \epsilon_{abc}$.

The sense of the considered representations is that upon excluding rotation matrices by classical equations of motion (that is, on-shell) these result in the same (half of) Regge action. Taking into account that in the Minkowsky case $\plS = ( \miS )^*$ we can write out the most general combination of $\plS$, $\miS$ which i) reduces to Regge action on-shell and ii) is real, as $S$ = $C \plS + C^* \miS$ where $C + C^* = 2$, that is $C = 1 + i \cdot \mbox{(real parameter)}$. At the same time, in the continuum theory the {\it Holst} action which generalizes the Cartan-Weyl form of the Einstein action \cite{Holst,Fat} is easily seen to have the form $(1 + i/\gamma) \plS_{\rm cont} + (1 - i/\gamma) \miS_{\rm cont}$ where $\pmS_{\rm cont}$ are (anti-)selfdual parts of the Cartan-Weyl continuum action, $\gamma$ is known as Barbero-Immirzi parameter \cite{Barb,Imm}. Therefore we can write $C = 1 + i / \gamma$ where the discrete analog of $\gamma$ is denoted by the same letter. Thus
\begin{equation}\label{S+S}
S = \left (1 + {i\over \gamma}\right )\plS + \left (1 - {i\over \gamma}\right )\miS.
\end{equation}

\noindent In the considered case of bisimplex $\pmS$ implies $\pmSSUtwbisim (\sfo )$ or $\pmSSOthbisim (\sfo )$.

Rewrite (\ref{SU2}), (\ref{SO3}) in more specialized notations. Denote the vertices of the $\sfo$ by $i = 0, 1, 2, 3, 4$. Denote a simplex by enumerating its vertices in round brackets. Then $\sfo = (01234), \sfoprime = (0^{\prime}1234)$ with the common 3-face $(1234)$ and vertices $0, 0^{\prime}$ identified. Let all the connections $\Omega$ on the 3-faces act from
(01234) to $(0^{\prime}1234)$, that is, if a 2-face tensor $v$ is defined in
(01234), then $\Omega v \OmT$ is defined in $(0^{\prime}1234)$. Denote a
3-face in the same way as the opposite vertex, $\Omega_i$ $\equiv$ $\Omega_{\sigma^3}$
where $\sigma^3$ = $(\{01234\}\setminus \{i\})$. Here \{\dots\} denote (sub)set, here
of the vertices 0, 1, 2, 3, 4. Denote a 2-face in the same way as the opposite edge,
$v_{(ik)}$ $\equiv$ $v_{\sigma^2}$, $R_{(ik)}$ $\equiv$ $R_{\sigma^2}$ where
$\sigma^2$ = $(\{01234\}\setminus \{ik\})$. It is convenient to define the variables
$v$, $R$ on the ordered pairs of vertices $ik$, $v_{ik}$ = $-v_{ki}$, $R_{ik}$ =
$\RT_{ki}$. (Then $v_{(ik)}$ is a one of the two values, $v_{ik}$ or $v_{ki}$,
$R_{(ik)}$ is $R_{ik}$ or $R_{ki}$). Then
\begin{equation}                                                                    %2
R_{ik} = \OmT_i\Omega_k.
\end{equation}

\noindent Evidently,
 \begin{equation}\label{Bianchi}                                                                   R_{ik}R_{kl} = R_{il}
\end{equation}

\noindent (these are the Bianchi identities \cite{Regge}
on the triangles with common edge $(\{01234\}\setminus \{ikl\})$). As independent
curvature matrices we can choose $R_{\alpha}$ $\equiv$ $R_{0\alpha}$ ($\alpha$,
$\beta$, $\gamma$, \dots = 1, 2, 3, 4), that is, the curvature on the 2-faces of the
tetrahedron (1234). With the shorthand $v_{\alpha}$ $\equiv$ $v_{0\alpha}$ the actions (\ref{SU2}), (\ref{SO3}) read
\begin{eqnarray}
\label{SU2-detail}%
\pmSSUtwbisim = \sumalpha \sqrt{\pmbvalpha^2} \Arcsin {\pmvalpha \circ \pmR_{\alpha} \over \sqrt{\pmbvalpha^2}} + \sumalphabeta \sqrt{\pmbvalphabeta^2} \Arcsin {\pmvalphabeta \circ ( \pmRT_{\alpha} \pmR_{\beta} ) \over \sqrt{\pmbvalphabeta^2}}, \\
\pmSSOthbisim = \sumalpha \frac{1}{2} \sqrt{\pmbvalpha^2} \Arcsin {\pmbvalpha * \pmR_{\alpha} \over \sqrt{\pmbvalpha^2}} + \sumalphabeta \frac{1}{2} \sqrt{\pmbvalphabeta^2} \Arcsin {\pmbvalphabeta * ( \pmRT_{\alpha} \pmR_{\beta} ) \over \sqrt{\pmbvalphabeta^2}}.
\end{eqnarray}

\noindent The matrix $R_{\alpha}$ solves the equations of motion for connections if it is the rotation by an angle $(2\pi - 2\alstwsfo )$ around a 2-face $(\{01234\}\setminus \{0 \alpha \})$ where $\alstwsfo$ is hyper-dihedral angle %between the 3-faces $(\{1234\}\setminus \{ \alpha \})$ and $(\{1234\})$
of the considered 4-simplex at this 2-face. Apart from the four $R_{\alpha}$s, one of the matrices $\Omega$, say $\Omega_0$, can be taken as fifth, purely gauge connection variable absorbing the rotations of the local frame.

Some non-standard feature of the considered representations is their nonperturbative nature. Matrices $\Omega, R$ cannot be considered as all of these close to unity because bisimplex is strongly curved manifold.

\vspace{-5mm}

\renewcommand{\baselinestretch}{1.2}

\section{Representation of arbitrary Regge action from representation of bisimplex action}\label{full-action}

\renewcommand{\baselinestretch}{1.5}

\vspace{-3mm}

Important is that according to formula (\ref{Sbisim}) we also have connection representation for the following combination,
\begin{equation}
\sumstwsubsfo \alstwsfo \Astwsfo,
\end{equation}

\noindent for the given 4-simplex $\sfo$. On the other hand, the same combinations appear in the Regge action for any collection of the 4-simplices,
\begin{eqnarray}\label{Regge-total}
\nonumber S & = & \sumstw \left ( 2\pi - \sumsfosupstw \alstwsfo \right ) \Astwsfo \\ \nonumber & = & 2\pi \sumstw \Astw - \sumsfo \sumsfosupstw \alstwsfo \Astwsfo \\ & = & \sumsfo \left [ {1\over 2}\Sbisimsfo + \sumstwsubsfo \left ({2\pi\over \Nstw} - \pi \right )\Astwsfo \right ],\label{S/2}%
\end{eqnarray}

\noindent where $\Nstw$ is the number of the 4-simplices meeting at $\stw$. Here $\Sbisimsfo$ can be substituted by expression (\ref{S+S}) where $\pmSSUtwbisim (\sfo )$ or $\pmSSOthbisim (\sfo )$ stand for $\pmS$. This gives some representations for the Regge action for an arbitrary simplicial complex. Usual Regge calculus implies independence of $\Astwsfo$ on $\sfo \supset \stw$: $\Astwsfo \equiv \Astw$. The unusual feature of the considered representations is their nonperturbative nature, as above for the bisimplex alone. Even in the continuum limit or near the flat background matrices $\Omega, R$ involved cannot be treated as close to unity.

Physical meaning of a matrix $R$ used in the considered representation is rotation by an angle $(2\pi - 2\alstwsfo )$ around certain 2-face $\stw$ where $\alstwsfo$ is hyperdihedral angle of the 4-simplex $\sfo$ on this face. There is no need in the explicit presence of the bisimplices in general simplicial complex. Sufficient is that when excluding rotations via the equations of motion from (\ref{Regge-total}) we get exactly Regge action for the general simplicial spacetime.

\vspace{-7mm}

\section{Application to path integral formulation}\label{pathint}

\vspace{-3mm}

Let us write out a discretized functional integral $\int \exp (iS) D q$, $q$ are field variables (some factors of the type of Jacobians could also be present). Suppose we are interested in the result of integration over connections as function of area tensors. Of course, different (components of) area tensors are not independent, but nothing prevent us from studying analytical properties in the extended region of varying these area tensors as if these were independent variables. This result splits into separate factors corresponding to integration over connection matrices $\Omsthsfo$ in the separate 4-simplices $\sfo$,
\begin{eqnarray}
& & \int \exp (iS) \prodsthsfo \D \Omsthsfo = \exp \left [ i \sumsfo \sumsthsubsfo \left ( \frac{2\pi}{\Nstw} - \pi \right ) \Astwsfo \right ] \nonumber \\ & & \cdot \prodsfo \int \exp (i\Sbisimsfo / 2) \prodsthsubsfo \D \Omsthsfo.
\end{eqnarray}

\noindent Each such factor is certain function of area tensors in the given 4-simplex proportional to (in the case of SO(3) $\!\times\!$ SO(3) representation)
\begin{eqnarray}\label{N}
& & \N (\{ \valpha \} , \{ \valphabeta \} ) = \int \exp \frac{i}{4} \left \{ \left ( 1 + \frac{i}{\gamma }\right ) \left [ \sumalpha \sqrt{\pbvalpha^2} \arcsin {\pbvalpha * \pR_{\alpha} \over \sqrt{\pbvalpha^2}} \right. \right. \nonumber \\ & & \left. \left. + \sumalphabeta \sqrt{\pbvalphabeta^2} \arcsin {\pbvalphabeta * ( \pRT_{\alpha} \pR_{\beta} ) \over \sqrt{\pbvalphabeta^2}} \right ] + \mbox{\rm complex conjugate} \right \} \prodalpha \D R_{\alpha},
\end{eqnarray}

\noindent after reducing '$\Arcsin$'s to the principal values '$\arcsin$'s and, probably, redefining $\pmbvalphabeta \to - \pmbvalphabeta$ as considered in Section \ref{branches}. We have taken into account that
\begin{equation}
\prodsthsubsfo \D \Omsthsfo = \prod^4_{i=0} \D \Omega_i = \D \Omega_0 \prodalpha \D R_{\alpha}
\end{equation}

\noindent and have divided by the volume of gauge group by omitting $\D \Omega_0$. Here
\begin{eqnarray}
\D R = \D \pR \D \mR = \frac{\sin^2 \left (\pphi /2 \right )}{ 4\pi^2 \pphi^2} \frac{\sin^2 \left (\mphi /2 \right )}{ 4\pi^2 \mphi^2} \d^3 \pbphi \d^3 \mbphi \nonumber \\ = \left (\frac{1}{ \sqrt{1 - \pbr^2}} - 1 \right ) \left (\frac{1}{ \sqrt{1 - \mbr^2}} - 1 \right ) \frac{\d^3 \pbr }{ 8\pi^2 \pbr^2} \frac{\d^3 \mbr }{ 8\pi^2 \mbr^2}.
\end{eqnarray}

\noindent where $\pmr^a = \epsilon^a{}_{bc} \pmR^{bc} / 2 = \left (\pmphi^a \sin \pmphi \right ) / \pmphi$, $\pmphi = \sqrt{\pmbphi^2}$, $\pmbphi = \bvarphi \mp i\bpsi$ is complex angle parameter (\ref{varphi-mp-ipsi}), here of the rotation $R$, and definition of the integration element for complex values, $\pbr = (\mbr )^*$, is $\d^3 \pbr \d^3 \mbr = 2^3 \d^3 \Re \pbr \d^3 \Im \pbr$.

Compare $\N (\{ \valpha \} , \{ \valphabeta \} )$ with the expression following from (\ref{N}) by taking in the curvatures $\pmR_{\alpha}$ in the exponential and in the measure $\D \pmR_{\alpha}$ their parts linear in $\pmbralpha$. (In this procedure, the SU(2) $\!\times\!$ SU(2) version of Eq. (\ref{N}) gives the same.) We get
\begin{eqnarray}\label{r}
\int \exp \frac{i}{4} \left \{ \left ( 1 + \frac{i}{\gamma }\right ) \left [ \sumalpha \pbvalpha \pbralpha + \sumalphabeta \pbvalphabeta (\pbrbeta - \pbralpha ) \right ] \right. \nonumber \\ \left. \phantom{\left ( \left [ \sumalpha \right ] \right )} + \mbox{\rm complex conjugate} \right \} \prodalpha \frac{\d^3 \pbralpha \d^3 \mbralpha}{(16 \pi^2)^2} \nonumber \\ = \prodalpha \frac{(64 \pi)^2 \gamma^6}{(1 + \gamma^2)^3} \deth \left ( \pbvalpha + \sumbeta \pbvbetaalpha \right ) \deth \left ( \mbvalpha + \sumbeta \mbvbetaalpha \right )
\end{eqnarray}

\noindent where \dfuns of complex values have sense as $\deth (\pbv) \deth (\mbv) = \deth (\pbv) \deth ((\pbv)^*) \equiv 2^{-3} \deth (\Re \pbv) \deth (\Im \pbv)$. That is, we pass from the nonlinear manifold SO(3,1) to its tangential hyperplane, so(3,1). Thus we get \dfuns expressing closeness of the surfaces of the tetrahedrons, that is, relations of the type of Gauss law which should hold identically on real physical system.

The (\ref{r}) is Fourier transform of 1 on hyperplane of so(3,1). In reality, we have Fourier transform on the nonlinear manifold SO(3,1). This should lead to smoothed and broadened \dfuns. To reveal the type of this broadening, consider integrals of the function of interest $\N (\{ \valpha \} , \{ \valphabeta \} )$ with products of the components $\pmvalpha^a$ (monomials),
\begin{equation}\label{moment}
\int \N \prod_{a, \alpha} \left ( \pvalpha^a \right )^{\pkalpha^a} \left ( \mvalpha^a \right )^{\mkalpha^a} \d \pvalpha^a \d \mvalpha^a,
\end{equation}

\noindent so-called moments. Let us use the calculational model with the function '$\arcsin$' in the action being linearized. To define (\ref{moment}), we first integrate $\exp (iS)$ over $\!\prod_{\alpha}\!$ $\!\d^3 \! \pbvalpha\!$ $\! \d^3 \! \mbvalpha\!$, then over $\prod_{\alpha} \D R_{\alpha}$. Upon first integration we get the product of (the derivatives of) the \dfuns
\begin{equation}
\prodalpha \deth \left ( \pR_{\alpha} - \pRT_{\alpha} \right ) \deth \left ( \mR_{\alpha} - \mRT_{\alpha} \right ).
\end{equation}

\noindent Occurrence of the support of this at zero $\pmbralpha$s somewhat justifies the adopted calculational model with linearized '$\arcsin$' as if the $\pmbralpha$s were small. So this model should be qualitatively correct. Subsequent integration over $\prod_\alpha \D R_{\alpha}$ gives finite answer. Note that if integration over $\d^6 \valphabeta$ (may be, with some product of the components of $\valphabeta$) were additionally inserted into the definition of moment, this integration would lead to singularity of the type of $\desi (0)$,
\begin{equation}
\deth ( \!\pRT_\beta \! \pR_\gamma \! - \! \pRT_\gamma \! \pR_\beta \! ) \deth ( \!\mRT_\beta \! \mR_\gamma \! - \! \mRT_\gamma \! \mR_\beta \! ) \prodalpha \deth \left ( \! \pR_{\alpha} \! - \! \pRT_{\alpha} \right ) \deth \left ( \! \mR_{\alpha} \! - \! \mRT_{\alpha} \right ).
\end{equation}

\noindent Finiteness of the integral of a function with any product of its arguments means that this function is decreasing faster than any inverse power of arguments. The simplest such function is exponentially decreasing one. This type of decreasing is most natural in the present case when the function of interest $\N$ is itself integral of exponent (and, by proper deformation of integration contours in complex plane, is expected to be representable as a priori combination of increasing and decreasing monotonic exponents). So the function $\N$ should exponentially decay at large $\valpha$s and fixed $\valphabeta$s. Thus, when considering the exact $\N$, \dfuns (\ref{r}) are expected to be broadened like decreasing exponent. (Vice versa, \dfun is itself a limiting case of the extremely rapidly decreasing exponent.)

Important is that since the different terms in the argument of delta, $\pmbvalpha \! + \! \sumbeta \!\!\!\! \pmbvbetaalpha$, enter exact expression (\ref{N}) in the different way, we expect that there should be suppression not only over such sum of these terms, but also over these terms separately at large their values.

On the other hand, defining integral over connections as function of area tensors from knowing its moments, i. e. integrals with products of area tensor components, is itself physically sensible way of defining conditionally convergent integral such as path integral. The moments have physical sense of the expectation values for the products of area tensor components in a theory with independent area tensors.

In more detail, $\N$ is exactly calculable for zero $\valphabeta$s when the $\N$ factorizes into functions of separate $\valpha$s,
\begin{eqnarray}
& & \N (\{ \valpha \}, \{ 0 \}) = \prodalpha \N_0 (\valpha ), \\ & & \label{N0SO(3)} \hspace{-15mm} \N_0 ( v ) = \int \exp \frac{i}{4} \left [ \left ( 1 + \frac{i}{\gamma} \right ) \sqrt{ \pbv^2 } \arcsin \frac{ \pbv * \pR }{ \sqrt{ \pbv^2 } } + \mbox{complex conjugate} \right ] \D R.
\end{eqnarray}

\noindent In the calculational model with linearized '$\arcsin$' we get for $\N_0$, in accordance with SO(4) = SO(3) $\!\times\!$ SO(3) (more accurately, plus the same times the element $-1$ of SO(4)), the square of the (analytically continued in proper way) suppression factor for the length $l$ in SO(3) gravity \cite{Kha4}. The latter was defined from knowing edge expectation values, i. e. its moments. It turns out to be proportional to $Ki_1 (l) / l$, $Ki_1 (l)$ is modified integral Bessel function. Now
\begin{equation}
\N_0 (v) = \left | \frac{Ki_1 \left ( \frac{1}{4} \sqrt{\left ( \frac{1}{\gamma} - i \right )^2 \pbv^2 } \right ) }{\frac{\pi}{2} \sqrt{\left ( \frac{1}{\gamma} - i \right )^2 \pbv^2 } } \right |^2, ~~~ Ki_1 (l) = \int\limits^{\pi / 2}_0 \exp \left ( - \frac{l}{\sin \varphi} \right ) \d \varphi .
\end{equation}

\noindent At $\pmbv^2 = - | \bv |^2$ (spacelike region) $\N_0 (v)$ behaves as $\exp ( - |\bv | / 2 )$. At $\pmbv^2 = | \bv |^2$ (timelike region) $\N_0 ( v )$ behaves as $\exp ( - | \bv | / (2 \gamma ) )$.

The above exponential suppression over areas can be illustrated by the model integral
\begin{equation}
\int\limits^{+\infty}_{-\infty} e^{i\sqrt{-v^2} \sh \psi}\d \psi = 2K_0(\sqrt{-v^2}), ~~~ K_0 (l) = \int\limits^{\pi / 2}_0 \exp \left ( - \frac{l}{\sin \varphi} \right ) \frac{ \d \varphi }{\sin \varphi } ,
\end{equation}

\noindent Here $\sqrt{-v^2}$ is modeling module of the spacelike area, $\psi$ is modeling Lorentz boost angle. This behaves as $\exp (-\sqrt{-v^2})$ at large $v^2$. Nonzero $\gamma^{-1}$ mixes spacelike and timelike area components and leads to exponential suppression also in the timelike region.

The $\N_0 ( v )$ with exact function '$\arcsin$' can be exactly computed and is exponentially suppressed at large $|\pmbv^2|$ as well \cite{Kha22}. Due to the oddness of '$\arcsin$', a moment (\ref{moment}) can be reduced to include integrations over new variables, which are areas $\sqrt{\pmbvalpha^2}$ with sign. Integration over these just give (the derivatives of) the \dfuns of both '$\arcsin$' functions in Eq. (\ref{N}) which being further integrated allows to define the moment.

In the case of SU(2) $\!\times\!$ SU(2) representation some more complicated expressions for $\N$ can be written explicitly in the form of absolutely convergent exponentially suppressed integrals. Suppose we define integral over connections as function of area tensors from knowing its moments, i. e. integrals with products of area tensor components. Then this integral can be obtained in simple way from the formal expression of the Euclidean version of this integral by deforming integration contours over $\D R \sim \d^3 \! \pbphi \, \d^3 \! \mbphi$ into complex plane in certain way \cite{Kha5}. Namely, we write $\pmbphi$ in the spherical coordinates relative to the direction of $\pmbn = \pmbphi / \pmphi$ corresponding to the lowest value of the exponential. The radial coordinate is then deformed as
\begin{equation}
\frac{1}{2} \pmphi \Rightarrow \frac{ \pi }{2} + i \pmeta , ~~~ - \infty < \pmeta < + \infty .
\end{equation}

\noindent The azimuthal angle of $\pmbphi$ is deformed as
\begin{equation}
\pmtheta \Rightarrow i \pmzeta , ~~~ 0 \leq \pmzeta < + \infty .
\end{equation}

\noindent The polar angle $\pmchi$ remains unchanged. For the SU(2) $\!\times\!$ SU(2) modification of (\ref{N0SO(3)}) and in the calculational model with linearized '$\arcsin$',
\begin{equation}
\N_0 ( v ) = \int \exp \frac{i}{2} \left [ \left ( 1 + \frac{i}{\gamma} \right ) \pv \circ \pR + \mbox{complex conjugate} \right ] \D R
\end{equation}

\noindent we write $\pmv \circ \pmR = \pmbv \pmbn \sin ( \pmphi / 2 )$ and
\begin{eqnarray}
& & \N_0 ( v ) = \int \exp \left (  - \plm \circ \pR - \mim \circ \mR \right ) \D R \nonumber \\ & & = (4 \pi^2)^{-2} \int \exp \left (  - \sqrt{\pbm^2 } \ch \peta \ch \pzeta - \sqrt{\mbm^2 } \ch \meta \ch \mzeta \right ) \nonumber \\ & & \label{N0K1} \cdot \ch^2 \!\peta \,\d \!\peta \,\d \ch \!\pzeta \,\d \!\pchi \ch^2 \!\meta \,\d \!\meta \,\d \ch \!\mzeta \,\d \!\mchi = \frac{K_1 (\sqrt{\pbm^2})}{\pi \sqrt{\pbm^2}} \frac{K_1 (\sqrt{\mbm^2})}{\pi \sqrt{\mbm^2}} \\ & & = \left | \frac{K_1 \left [ \frac{1}{2} \sqrt{\left ( \frac{1}{\gamma} - i \right )^2 \pbv^2} \right ]}{ \frac{\pi}{2} \sqrt{\left ( \frac{1}{\gamma} - i \right )^2 \pbv^2} } \right |^2 , ~~~ K_1 (l) = \int\limits^{\pi / 2}_0 \exp \left ( - \frac{l}{\sin \varphi} \right ) \frac{ \d \varphi }{\sin^2 \varphi } ,
\end{eqnarray}

\noindent as the result of continuation to
\begin{equation}
\pmm = - \frac{i}{2} \left ( 1 \pm \frac{i}{\gamma } \right ) \pmv .
\end{equation}

\noindent There analytically continued unit vector of the rotation $\pmbn$ providing the lowest value in the exponential is $\pmbv (\pmbv^2)^{-1/2}$ with certain sign. Generally it can be written as
\begin{equation}
\pmbn^{(0)} = - i (1 \pm i / \gamma ) [- (1 \pm i / \gamma )^2 \pmbv^2 ]^{-1/2} \pmbv .
\end{equation}

\noindent In the physical case, $\Im \pmbv^2 = 0$, this is
\begin{equation}
\frac{\pm \pmbv}{\sqrt{(1 \mp i0) \pmbv^2}} = \frac{- i \pmbv}{\sqrt{ - (1 \pm i0) \pmbv^2}} = \left \{ \begin{array}{rcl} -i \pmbv ( - \pmbv^2)^{- 1/2} & \mbox{at} & \pmbv^2 < 0 , \\ \pm \pmbv ( \pmbv^2 )^{- 1/2 } & \mbox{at} & \pmbv^2 > 0
\end{array} \right .
\end{equation}

\noindent Relative to $\pmbn^{(0)}$, the $\pmbn$ is parameterized  in the spherical coordinates in usual way,
\begin{equation}
\pmbn = \pmbn^{(0)} \ch \pmeta + i (\sh \pmzeta) ( \pmbe_1 \cos \pmchi + \pmbe_2 \sin \pmchi ) .
\end{equation}

\noindent Here $\pmbe_1, \pmbe_2$ are the two unit vectors constituting with $\pmbv (\pmbv^2 )^{- 1/2}$ an orthonormal triple.

Generally we have in the exponential also a scalar part $\pmm_0$ in $\pmm$,
\begin{equation}
\pmm = \frac{1}{2} \pmbm \pmbSig + \frac{1}{2} \pmm_0 \cdot 1 ,
\end{equation}

\noindent so that
\begin{equation}
\pmm \circ \pmR = \pmm_0 \cos \frac{ \!\!\! \pmphi}{2} + \pmbm \pmbn \sin \frac{ \!\!\! \pmphi}{2} .
\end{equation}

\noindent So we have combination of $\cos ( \pmphi / 2 )$ and $\sin ( \pmphi / 2 )$ in the exponential reducing to $\sin ( \pmphi / 2 + \pmbeta )$. Above deformation of integration contours over $\D R$ is applied to the shifted radial variable,
\begin{equation}
\frac{1}{2} \pmphi \Rightarrow \frac{ \pi }{2} + i \pmeta - \pmbeta .
\end{equation}

\noindent Calculation shows that the result (\ref{N0K1}) gets naturally generalized by replacing $\pmbm^2 \Rightarrow 2 \pmm \circ \pmm = \pmbm^2 + \pmm_0^2$. Using this, consider the 4-simplex with one of the edge, say $(40)$, small so that we can take $v_{23} = 0, v_1 \equiv v_{01} = - v_{41}$, \dots cycle perm (1,2,3) \dots . In the calculational model with linearized '$\arcsin$' the $\N$ takes the form
\begin{eqnarray}
& & \N_0 ( v ) = \int \exp \frac{i}{2} \left [ \left ( 1 + \frac{i}{\gamma} \right ) ( \pv_1 \circ \pR_1 + \pv_2 \circ \pR_2 + \pv_3 \circ \pR_3 + \ptilv_4 \circ \pR_4 ) \right. \nonumber \\ & & \left. \phantom{\prodalpha } + \mbox{complex conjugate} \right ] \prodalpha \D R_{\alpha}
\end{eqnarray}

\noindent where $\pmtilv_4 \equiv \pmv_4 - \pmR_1 \pmv_1 - \pmR_2 \pmv_2 - \pmR_3 \pmv_3$. Applying the above considered deformation of integration contours we get
\begin{eqnarray}
& & \N = \pcalN \mcalN , \\ & & \label{N40zero} \pmcalN = \int \frac{K_1 \left [ \frac{1}{2} \sqrt{\left ( \frac{1}{\gamma} \mp i \right )^2 {\rm tr} ( \pmtilv^{\rm T}_4 \pmtilv_4 ) } \right ]}{ \frac{\pi}{2} \sqrt{\left ( \frac{1}{\gamma} \mp i \right )^2 {\rm tr} ( \pmtilv^{\rm T}_4 \pmtilv_4 ) } } \nonumber \\ & & \cdot \prod^3_{\alpha = 1} \exp \left [ - \frac{1}{2} \sqrt{\left ( \frac{i}{\gamma } \mp i \right )^2 \pmbvalpha^2} \ch \pmeta_{\alpha} \ch \pmzeta_{\alpha} \right ] \ch^2 \pmeta_{\alpha} \d \pmeta_{\alpha} \d \ch \pmzeta_{\alpha} \d \pmchi_{\alpha}
\end{eqnarray}

\noindent where $\pmtilv_4$ depends on $\pmeta_\alpha , \pmzeta_\alpha , \pmchi_\alpha$ through $R_\alpha$ parameterized by these as above considered,
\begin{eqnarray}
& & \pmR_\alpha = - i \sh \pmeta_\alpha + \pmbSig \pmbn_\alpha \ch \pmeta_\alpha, \\ & & \pmbn_\alpha = \frac{-i \left ( 1 \pm \frac{i}{\gamma} \right ) \pmbv_\alpha }{ \sqrt{ - \left ( 1 \pm \frac{i}{\gamma} \right )^2 \pmbv^2_\alpha }} \ch \pmzeta_\alpha + i (\sh \pmzeta_\alpha ) (\pmbe_{1 \alpha }\cos \pmchi_\alpha + \pmbe_{2 \alpha } \sin \pmchi_\alpha ) .
\end{eqnarray}

\noindent Note that everywhere $\Re \sqrt{z} \geq 0$ for our choice of the branch of function $\sqrt{z}$ with the cut along negative real half-axis in the complex plane of $z$ such that $\sqrt{1} = 1$.

The Eq. (\ref{N40zero}) illustrates the above mentioned broadening the $\delta$-functions expressing closure conditions for the tetrahedrons. In this simplified configuration with zero edge $(04)$ this condition is $\pmv_1 +\pmv_2 + \pmv_3 - \pmv_4 = 0$. In the Eq. (\ref{N40zero}) we have exponentially dumped $K_1$ instead of \dfun and, besides, $\pmv_\alpha$s enter its argument not exactly in this combination, but with $\pmR_\alpha$ matrix coefficients which are in no way close to unity. Besides, there are also decreasing exponents showing suppression over $\pmvalpha$s separately.

In the case of SO(3) $\!\times\!$ SO(3) representation the unit vector $\pmbn$ enters exponential bilinearly, see (\ref{Omega-SO3}), and explicitly reducing $\N$ to the absolutely convergent exponentially suppressed integrals is a more difficult problem.

To see what does suppression of areas mean for suppression of the lengths, introduce a tetrad taken, say, at a vertex 4, $\lfoze^a, \lfoon^a, \lfotw^a, \lfoth^a$. The $l_{ik}$ means vector directed from the vertex $i$ to $k$. Then $v_1 = [ \lfotw , \lfoth ], v_{12} = [\lfoze, \lfoth]$, \dots cycle perm (1,2,3) \dots . The other four bivectors are defined from closure of the appropriate 3-simplices, $v_{41} = [ \lfoth - \lfoon , \lfotw - \lfoon ]$, \dots cycle perm (1,2,3) \dots , $v_4 = [ \lfotw - \lfoth , \lfoon - \lfoth ]$. Standard quantum gravity settings suggest choice of one of these vectors $\lfoze^a$ as lapse-shift one and Schwinger "time gauge" \cite{Schw} for three others, $\lfoalpha^0 = 0, \alpha = 1, 2, 3$. Then $2 \pmbv_1 = \pm i \blfotw \times \blfoth \equiv 2 \bv_1 , 2 \pmbv_{12} = \pm i \blfoon \times \blfoth + l^0_{40} \blfoth$, \dots cycle perm (1,2,3) \dots . Suppression of large $\bvalpha$s means suppression of large $\blfoalpha$s since small $\bvalpha$s mean small $\blfoalpha$s with exception of the region where $| \bv_1 \times \bv_2 \cdot \bv_3 | \to 0$. The latter region can be suppressed by the positive power of $-\det \| g_{\lambda\mu} \| = 8 ( \lfoze^0 )^2 | \bv_1 \times \bv_2 \cdot \bv_3 |$ in (the edge vector part of) the measure. The measure is usually considered as determinable up to a power of $-\det \| g_{\lambda\mu} \|$. (In fact, the first principles allow to fix the measure up to a power of $-\det \| g_{\lambda\mu} \|$ \cite{Mis,DeW}.)

\vspace{-7mm}

\section{Specifying analytical form of the representation}\label{branches}

\vspace{-3mm}

Generally, when defining quantum amplitude of transition between the two three-dimensional geometries in the simplicial framework, we should sum over all simplicial four-dimensional geometries interpolating between these two as over paths in the path integral. Of these simplicial geometries those ones with arbitrarily small edge lengths can arbitrarily accurately approximate the usual smooth geometries. So we do not lose essential continuum degrees of freedom and such properties as the diffeomorphism symmetry typical for the continuum should be restored in exact analysis. In practical calculations summation over all simplicial structures is recently technically unachievable, and we need to specify simplicial structure to work with.

Here we express considered in the present paper representations for action in terms of principal values of '$\arcsin$' function. Assuming simple periodic simplicial structure as basic example, the whole region of variation of edge lengths gets divided into different regions in which the action is combination of the '$\arcsin$' functions together with terms which are constants different in the different regions. These terms are important for possible integration over edge lengths in the path integral. These regions are particularly specified by performing triangulation consistent with the standard canonical quantization scheme using Schwinger time gauge.

Consider the simplest periodic simplicial structure used in \cite{RocWil} when the spacetime is divided into 4-cubes. To each vertex of this 4-cubic lattice one of sixteen 4-cubes containing this vertex is assigned located in the directions of the four main cubic axes defined as positive direction. The set of links of the simplicial complex is union over vertices of the sets of all the edges and 2-face, 3-face and 4-cube diagonals of the corresponding 4-cube emanating from the vertex. Each 4-cube is thereby divided into 24 4-simplices. Each 4-simplex contains four edges directed along all four main 4-cube axes. Further we consider one of such 4-simplices.

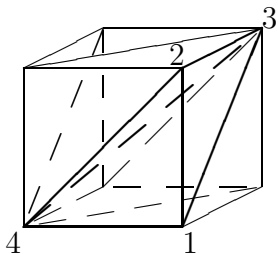
\begin{figure}[h]
\setlength{\unitlength}{1pt}
\begin{picture}(100,90)(-160,0)
\put (0,0){\line(0,1){60}}
\put (0,60){\line(1,0){60}}
\put (0,60){\line(2,1){30}}
\put (60,0){\line(2,1){30}}
\put (30,75){\line(1,0){60}}
\put (90,15){\line(0,1){60}}

\put (0,60){\line(6,1){90}}

\multiput (2,5)(8,20){4}{\line(2,5){4}}
\multiput (30,15)(0,17){4}{\line(0,1){8}}
\multiput (4,2)(16,8){2}{\line(2,1){10}}
\multiput (34,15)(20,0){3}{\line(1,0){10}}
\multiput (30,15)(16,16){4}{\line(1,1){10}}
\multiput (6,1)(18,3){5}{\line(6,1){10}}

\thicklines
\put (0,0){\line(1,0){60}}
\put (60,0){\line(0,1){60}}
\put (60,60){\line(2,1){30}}

\multiput (0,0)(18,15){5}{\line(6,5){10}}

\put (0,0){\line(1,1){60}}
\put (60,0){\line(2,5){30}}
\put (-7,-10){$4$}
\put (60,-10){$1$}
\put (55,61){$2$}
\put (90,75){$3$}
\end{picture}
\caption{The 3-cube divided into six 3-simplices.}\label{3cube}
\end{figure}

According to the notations of the end of Section \ref{pathint}, we have a tetrad taken at a vertex 4 of a 4-simplex of the considered simplicial complex, $\lfoze^a, \lfoon^a, \lfotw^a, \lfoth^a$. Standard quantum gravity settings suggest choice of one of these vectors $\lfoze$ as lapse-shift one and Schwinger time gauge for other three, $\lfoalpha^0 = 0, \alpha = 1, 2, 3$ (that is, $\lfoalpha, \alpha = 1, 2, 3$ are spacelike). Regarding lapse-shift as parameter for choice, we are free to choose $\lfoze$ i) timelike and ii) small (by the value of components) compared to typical scale of $| \blfoalpha |, \alpha = 1, 2, 3,$ so that $( \lfoalpha - \lfoze )^2$ be spacelike as $\lfoalpha$ are, $\alpha = 1, 2, 3$. Thus, the only timelike link in the simplex $(01234)$ is $(04)$, others are spacelike ones. The lapse-shift $(04)$ and analogous ones in other 4-simplices are supposed to be directed along the same one of the four main 4-cube axes. The other three main cubic axes define passing through the vertex 0 or 4 the three-dimensional section, see fig.\ref{3cube}.
This section is the three-dimensional version of the considered simplicial complex. There we have denoted by 3-simplex $(4123)$ that one being projection of the 4-simplex $(01234)$ onto the 3-section. (More accurately, some of the vertices 4, 1, 2, 3 in the notation $(4123)$ might be substituted by their projections $4^{\prime} = 0, 1^{\prime}, 2^{\prime}, 3^{\prime}$ onto the 3-section, that is, $(1^{\prime}1), (2^{\prime}2), (3^{\prime}3)$ are the lapse-shift edges.) Each such 3-simplex in the unperturbed flat symmetrical (w. r. t. the permutations of $(1,2,3)$) uniform such three-dimensional lattice is defined by three basic vectors, $\be_1 \! \equiv \, \stackrel{\longrightarrow}{41}$, $\be_2 \! \equiv \, \stackrel{\longrightarrow}{12}$, $\be_3 \! \equiv \, \stackrel{\longrightarrow}{23}$ (for the particular notations of the fig. \ref{3cube}). Up to an overall scale factor, these can be regarded as normalized, $\be^2_1 = \be^2_2 = \be^2_3 = 1$, but generally nonorthogonal, $\be_1 \be_2 = \be_2 \be_3 = \be_3 \be_1 \equiv \lambda $ (all three directions are treated symmetrically). The values of the dihedral angles in the 3-simplex are
\begin{eqnarray}\label{alpha3d}
& & \hspace{-15mm} \alpha_{2(43)1} = \pi / 3, ~~ \alpha_{1(42)3} = \alpha_{4(31)2} = \pi / 2, ~~ \alpha_{4(23)1} = \alpha_{3(41)2}, ~~ \alpha_{4(12)3} = \pi - 2 \alpha_{4(23)1},  \nonumber \\ & & \cos \alpha_{4(23)1} = \sqrt{(1 + 2 \lambda ) / (2 + 2 \lambda ) }.
\end{eqnarray}

\noindent It is known that as the minimal dihedral angle of a simplex is better bounded away from zero, the conditions for closeness between the continuum and discrete manifolds are better \cite{Cheeger}. It is seen that the minimal dihedral angle is largest at $\lambda = - 1 / 3$, when this angle is $\pi / 3$ (the 3-cubic lattice is shrunk along the body diagonal, so that we have rather parallelopipeds than cubes).

Above specifying simplicial geometry we define ranges for possible values of the dihedral angles in the typical 4-simplex. Denote by $\alpha_{0(123)4}$ the angle $\alstwsfo$ on $\stw = (123)$ in $\sfo = (01234)$ and analogously for others. It is not difficult to conclude that the angles on certain triangles are as follows
\begin{eqnarray}\label{pi/2+ieta}
& & \left. \begin{array}{c} \alpha_{0(234)1} = \frac{\pi}{2} + i \eta_{0(234)1} \\ \alpha_{1(023)4} = \frac{\pi}{2} + i \eta_{1(023)4}
\end{array} \right \} , \dots \mbox{2 cycle perm (1,2,3)} \dots \\ & &
\alpha_{0(123)4} = i \eta_{0(123)4}
\end{eqnarray}

\noindent The $\eta$ is everywhere real. Knowing possible ranges for the angles, it is not difficult to write identities relating angles to the values of '$\arcsin$' which can arise in the SO(3) $\!\times\!$ SO(3) representation (in the absence of torsion),
\begin{eqnarray}\label{0(234)1}
& & \left. \begin{array}{c} \alpha_{0(234)1} = \frac{\pi}{2} + \frac{1}{2} \arcsin \sin [2 \pi - 2 \alpha_{0(234)1} ] \\ \alpha_{1(023)4} = \frac{\pi}{2} + \frac{1}{2} \arcsin \sin [2 \pi - 2 \alpha_{1(023)4} ]
\end{array} \right \} , \dots \mbox{2 cycle perm (1,2,3)} \dots , \\ & & \label{0(123)4}
\alpha_{0(123)4} = - \frac{1}{2} \arcsin \sin [2 \pi - 2 \alpha_{0(123)4} ] , \\ & & \label{Eucl-angles} \alpha_{2(014)3} = \frac{\pi}{2} + \frac{1}{2} \arcsin \sin [2 \pi - 2 \alpha_{2(014)3} ] , \dots \mbox{2 cycle perm (1,2,3)} \dots .
\end{eqnarray}

\noindent The angles (\ref{Eucl-angles}) are real ones on the triangles $(014), (024), (034)$. In the three-di\-men\-si\-o\-nal section, these rotations correspond to the rotations by the usual Euclidean angles around the edges. The identity (\ref{Eucl-angles}) is valid in the region
\begin{equation}
\pi / 4 < \alpha_{1(024)3} < 3 \pi / 4 , \dots \mbox{2 cycle perm (1,2,3)} \dots .
\end{equation}

\noindent In the neighborhood of the point in the configuration superspace where the lapse-shift is orthogonal to the flat three-dimensional section, the considered hyperdihedral angles are close to the corresponding dihedral angles in the three-dimensional section (\ref{alpha3d}), $\alpha_{2(014)3} \approx \alpha_{2(14)3} , \dots $. Neighborhood of the pseudo-cubic (shrunk along the main cube diagonal) three-dimensional section at the parameter $\lambda = - 1 / 3$ and the dihedral angles $\pi / 3$ and $\pi / 2$ just fall into the region $( \pi / 4 , 3 \pi / 4 )$.

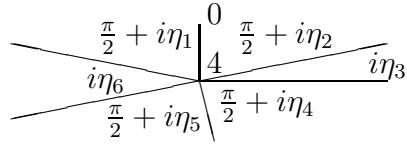
\begin{figure}[h]
\setlength{\unitlength}{0.5mm}
\begin{picture}(100,30)(-80,40)
\put (60,50){\line(0,1){15}}
\put (60,50){\line(1,-4){4}}
\put (60,50){\line(5,0){50}}
\put (60,50){\line(5,1){50}}
\put (60,50){\line(-5,1){50}}
\put (60,50){\line(-5,-1){50}}
\put (62,65){$0$}
\put (62,52){$4$}
\put (70,60){${\pi \over 2} + i\eta_2$}
\put (65,42){${\pi \over 2} + i\eta_4$}
\put (35,38){${\pi \over 2} + i\eta_5$}
\put (33,60){${\pi \over 2} + i\eta_1$}
\put (105,52){$i\eta_3$}
\put (30,48){$i\eta_6$}
\end{picture}
\caption{The pattern of six dihedral angles meeting at a triangle.}\label{angles}
\end{figure}

Now we can form the defect angles using the values of the angles (\ref{0(234)1}), (\ref{0(123)4}), (\ref{Eucl-angles}) in the considered and neighboring 4-simplices. Each angle (\ref{0(234)1}) is on the triangle on which there are three else angles of such type from neighboring 4-simplices, and, for the given simplicial structure, there are also two or null angles of the type of (\ref{0(123)4}), fig.\ref{angles}.
As a result, the constant real (proportional to $\pi$) part in the resulting defect angle is $2 \pi - 4 \cdot (\pi / 2) = 0$. Each angle (\ref{Eucl-angles}) is on the triangle on which there are three or five else angles of such type from neighboring 4-simplices (for the given simplicial structure). The constant real (proportional to $\pi$) part in the resulting defect angle is $2 \pi - N_{\sigma^2} \pi / 2 = 0$, $N_{\sigma^2}$ being the number of the 4-simplices sharing this triangle $\stw$. So there is part of the defect angle $\pi / N_{\stw} - \pi / 4$ per 4-simplex and per each part of the action, self- and antiselfdual one.

In thus defined region of edge lengths in the configuration superspace the action attributed to the given 4-simplex $(01234)$ takes the form
\begin{eqnarray}
& & \pmSSOth = \left ( \frac{ \pi }{ N_{(041)}} - \frac{ \pi }{ 4 } \right ) \sqrt{ \pmbv^2_{23} } - \frac{1}{4} \sqrt{ \pmbv^2_{23} } \arcsin \frac{\pmbv_{23} * \pmR_{23}}{ \sqrt{ \pmbv^2_{23} } } \nonumber \\ & & - \frac{1}{4} \sqrt{ \pmbv^2_{14} } \arcsin \frac{\pmbv_{14} * \pmR_{14}}{ \sqrt{ \pmbv^2_{14} } } - \frac{1}{4} \sqrt{ \pmbv^2_{01} } \arcsin \frac{\pmbv_{01} * \pmR_{01}}{ \sqrt{ \pmbv^2_{01} } } \nonumber \\ & & + \mbox{2 cycle perm (1,2,3)} + \frac{1}{4} \sqrt{ \pmbv^2_{04} } \arcsin \frac{\pmbv_{04} * \pmR_{04}}{ \sqrt{ \pmbv^2_{04} } } .
\end{eqnarray}

\noindent The number of the 4-simplices sharing the triangles, e. g. for the 4-simplex $(01234)$ built of the tetrahedron $(1234)$ of fig.\ref{3cube} and of lapse-shift $(40)$ is $N_{(041)}$ = 6 = $N_{(043)}$, $N_{(042)}$ = 4.

Relevant to the SU(2) $\!\times\!$ SU(2) representation identities for all the considered types of the angles can be written in the form
\begin{equation}
\alpha = \frac{\pi}{2} + \left [ \frac{\pi}{2} - \arcsin \sin (\pi - \alpha ) \right ] {\rm sgn} \Re \left ( \alpha - \frac{\pi}{2} \right ) .
\end{equation}

\noindent The natural region $0 < \Re \alpha < \pi$ is implied. For the angle of the type (\ref{0(123)4}) this gives
\begin{equation}
\alpha_{0(123)4} = \arcsin \sin (\pi - \alpha_{0(123)4} )
\end{equation}

\noindent like (\ref{0(123)4}) without the constant terms; in others the constant terms $\pi / 2$ remain as in the expressions for these (\ref{0(234)1}), (\ref{Eucl-angles}) as well. Therefore we get analogously
\begin{eqnarray}\label{SSU2}
& & \hspace{-8mm} \pmSSUtw = \left ( \frac{ \pi }{ N_{(041)}} - \frac{ \pi }{ 4 } \right ) \sqrt{ \pmbv^2_{23} } - \frac{1}{2} \sqrt{ \pmbv^2_{23} } \left [ \frac{\pi}{2} - \arcsin \frac{\pmv_{23} \circ \pmR_{23}}{ \sqrt{ \pmbv^2_{23} } } \right ] {\rm sgn} \Re \left ( \alpha_{23} - \frac{\pi}{2} \right ) \nonumber \\ & & - \frac{1}{2} \sqrt{ \pmbv^2_{14} } \left [ \frac{\pi}{2} - \arcsin \frac{\pmv_{14} \circ \pmR_{14}}{ \sqrt{ \pmbv^2_{14} } } \right ] {\rm sgn} \Re \left ( \alpha_{14} - \frac{\pi}{2} \right ) \nonumber \\ & & - \frac{1}{2} \sqrt{ \pmbv^2_{01} } \left [ \frac{\pi}{2} - \arcsin \frac{\pmv_{01} \circ \pmR_{01}}{ \sqrt{ \pmbv^2_{01} } } \right ] {\rm sgn} \Re \left ( \alpha_{01} - \frac{\pi}{2} \right ) \nonumber \\ & & + \mbox{2 cycle perm (1,2,3)} - \frac{1}{2} \sqrt{ \pmbv^2_{04} } \arcsin \frac{\pmv_{04} \circ \pmR_{04}}{ \sqrt{ \pmbv^2_{04} } } .
\end{eqnarray}

\noindent Here $\alpha_{23}$ is a shorthand for $\alpha_{2(014)3}$ etc. The $\alpha_{ik}$ in the RHS are implied to be functions of the edge lengths, and the whole region of variation of the edge lengths is divided into sectors in which sign functions keep their values. Some new feature of the SU(2) $\!\times\!$ SU(2) representation is ambiguity of the signs ${\rm sgn} \Re ( \alpha - \pi / 2 )$ in $\pmSSUtw$ for the angles (\ref{pi/2+ieta}) of the type $\pi / 2 + i \eta$. Ambiguity arises because $\arcsin z$ is here on the cut $\Im z = 0, z^2 > 1$, where it undergoes discontinuity. A way to fix these signs consistently might be to add to the lengths some infinitely small imaginary parts $\pm i 0$.

As far as certain expressions for area tensors in terms of edge vectors are not written yet, we can redefine area tensors in (\ref{SSU2}) including overall sign at '$\arcsin$' into their definition, e. g. $v_{23} {\rm sgn} \Re ( \alpha_{23} - \pi / 2 )$ $\Rightarrow$ $v_{23}$ etc. The same can be done in $\pmSSOth$, and, as far as the integration over connections is concerned, we deal with the standard forms of the action used in Section \ref{pathint} with all the '$\arcsin$'s entering with positive sign. The piecewise constant terms added to '$\arcsin$'s, different in the different sectors of edge length variation, lead to combination of areas added to action and should be taken into account when considering path integration over lengths. These terms reflect nonperturbative nature of the considered representations.

\vspace{-7mm}

\section{Conclusion}

\vspace{-3mm}

An attractive feature of the considered representations is that these allow to deal with comparatively simple simplicial gravity action instead of that one in terms of edge lengths only, with complicated trigonometric expressions. This is achieved by introducing additional rotational variables, but the dependence on these is simple and splits over separate terms in the action referred to separate 4-simplices.

In the path integral formalism, this leads to additional integrations over rotations, but these factorize over separate 4-simplices into ordinary integrals. Upon performing fixed finite number of integrations, we reduce the Minkowsky path integral to the form of absolutely convergent (as in the Euclidean version) exponentially suppressed (at large areas/lengths) integrals. This means suppression of the physical amplitudes with large areas/lengths (in Plank scale) and is important for the consistency of the simplicial minisuperspace system.

Qualitatively, this consideration can be repeated for the representation with the usual connections relating neighboring local frames of the 4-simplices, and the analogous conclusions concerning suppressing large areas/lengthgs can be made. To estimate the latter exactly, we should, strictly speaking, make arbitrarily large number of integrations over connections entering definition of path integral. Besides, passing to the variables with more clear physical sense, independent curvatures, is achieved via arbitrarily lengthy expressions which should express other curvatures in terms of independent ones (that is, resolve Bianchi identities).

Thus, simplifying points in the representation of the present paper are i) fixed finite number of additional rotational variables referred to separate 4-simplices; ii) the action is simple in terms of independent rotations since Bianchi identities are simple, (\ref{Bianchi}).

The present work was supported in part by the Russian Foundation for Basic Research
through Grants No. 08-02-00960-a and No. 09-01-00142-a.

\end{document}